\documentstyle[prl,aps,epsf]{revtex}
\twocolumn

\def\eps{\epsilon}
\def\fig#1{Fig.~\ref{#1}}
\def\siml{\,\hbox{\kern.1em \lower.6ex \hbox{$\sim$} \kern-1.12em
          \raise.6ex \hbox{$<$} }}
\def\simg{\,\hbox{\kern.1em \lower.6ex \hbox{$\sim$} \kern-1.12em
          \raise.6ex \hbox{$>$} }}
\newcommand{\be}{\begin{equation}}
\newcommand{\ee}{\end{equation}}
\newcommand{\Figurebb}[9]{
  \begin{figure}[H]\begin{center}
  \leavevmode
  \epsfysize=#7cm
  \epsfbox[#2 #3 #4 #5]{#6}
  \par
  \parbox{#8cm}{
  \caption[figure]{\renewcommand{\baselinestretch}{0.8} \small
                   \hspace{-0.3truecm}#9}\label{#1}}
  \end{center}
  \end{figure}
}

\begin{document}

\draft
\title{Semiclassical analysis of the lowest-order multipole deformations\\
of simple metal clusters}
\author{V. V. Pashkevich$^a$, P. Meier$^b$, M. Brack$^b$, and
        A. V. Unzhakova$^a$}
\address{$^a$Bogolyubov Laboratory of Theoretical Physics, JINR,
         141980 Dubna, Russia\\
         $^b$Institut f\"ur Theoretische Physik, Universit\"at Regensburg,
         D-93040 Regensburg, Germany}
\date{\today}
\maketitle

\vspace*{-0.65cm}

\begin{abstract}
We use a perturbative semiclassical trace formula to calculate the three 
lowest-order multipole (quadrupole $\eps_2$, octupole $\eps_3$, and 
hexadecapole $\eps_4$) deformations of simple metal clusters with $90 
\le N \le 550$ atoms in their ground states. The self-consistent mean 
field of the valence electrons is modeled by an axially deformed cavity 
and the oscillating part of the total energy is calculated semiclassically 
using the shortest periodic orbits. The average energy is obtained from a 
liquid-drop model adjusted to the empirical bulk and surface properties of 
the sodium metal. We obtain good qualitative agreement with the results of 
quantum-mechanical calculations using Strutinsky's shell-correction method.
\end{abstract}

\pacs{PACS numbers: 03.65.Sq, 05.30.Fk, 31.15.Ew, 71.10.Ca}

\vspace*{-0.2cm}

Free clusters made of simple metal atoms exhibit a pronounced electronic
shell structure \cite{knigt,deher,mattr}. Although the detailed experimental
information obtained, e.g., from photo-excitation measurements can only
be understood if the ionic structure is taken into account \cite{kuemm},
the qualitative features of the electronic shell structure can be well 
described, for not too small systems, by phenomenological deformed 
shell-model potentials \cite{clem,reim1,frpa}. Self-consistent density 
functional calculations in the framework of a deformed jellium model 
\cite{tom} have revealed that the cluster ground-state shapes can be well 
characterized in terms of the lowest three multipole orders $\eps_2$ 
(quadrupole), $\eps_3$ (octupole), and $\eps_4$ (hexadecapole). Since such 
self-consistent calculations are quite time consuming computationally, it 
is often more efficient to resort to simpler methods, such as the 
shell-correction method introduced by Strutinsky in nuclear physics 
\cite{strut}, in particular, if more shape degrees of freedom are to be 
investigated \cite{frpa}.

An even more economical approach is the semiclassical periodic orbit
theory (POT) (see, e.g., Ref.\ \cite{book} for a general introduction),
in which quantum oscillations in the level density or other observables 
can be described in terms of the leading shortest periodic orbits of the 
corresponding classical system through so-called trace formulae 
\cite{gutz,bablo}. This method has been used for quadrupole-deformed 
clusters in a Nilsson-type model \cite{reim2} and, more recently, using 
cavities with axial $\eps_2$, $\eps_3$, and $\eps_4$ deformations 
\cite{erice,peter}. The approximation of the self-consistent mean field of 
the valence electrons by a cavity with reflecting walls has received strong 
support from the quantitative explanation \cite{erice} of the experimental 
magic numbers found in connection with the electronic supershells 
\cite{supsh} in terms of the trace formula of the spherical cavity 
\cite{bablo}. The validity of the cavity model has also been confirmed by 
calculations with more realistic Woods-Saxon type potentials \cite{nish} 
and by selfconsistent Kohn-Sham calculations \cite{genz} in the spherical
jellium model.

In Ref.\ \cite{peter}, a perturbative trace formula derived by Creagh 
\cite{creag} has been used for axially deformed cavities with small 
multipole deformations $\eps_2$, $\eps_3$, or $\eps_4$ and found to 
reproduce the quantum-mechanical results very well for moderate values 
of the deformation parameters. The main virtue of this approach 
is that the perturbed orbits need not be known. The trace formula is 
given in terms of the periodic orbits of the unperturbed system
(i.e., here, the spherical cavity), their unperturbed amplitudes being 
multiplied by a modulation factor that contains the perturbations of 
their action integrals. The periodic orbits of the spherical 
cavity were discussed first by Balian and Bloch \cite{bablo} and
characterized by their winding number $t$ (number of revolutions around 
the centre) and the number $p$ of reflections from the boundary. 

In the present work we extend the approach developed in Ref.\ 
\cite{peter}. The deformed mean field of the electrons is param\-etrized 
by an axial cavity whose surface is described in terms of the Legendre 
polynomials $P_j$ by
\be
R(\theta) = R\,\Big(1+\eps_0+\sum_{j=2}^4\eps_j P_j(\cos\theta)\Big)\,,
\label{multipol2}
\ee
where $R=r_vN^{1/3}$ is the radius of the spherical cavity and $\eps_0$
is used to conserve its volume. In contrast to Ref.\ \cite{peter},
we include here simultaneously all three deformation parameters
$(def)=(\eps_2,\eps_3,\eps_4)$ and minimize for each cluster size $N$ 
the total energy
\be
E_{tot}(N,def) = E_{LDM}(N,def) + \delta E(N,def)\,,
\label{etot}
\ee
in order to determine its ground-state deformation. Here $E_{LDM}$ is 
the average total energy obtained in the liquid drop model (LDM) and 
$\delta E$ is the shell-correction energy. For the LDM model parameters 
and the radius constant $r_v$ we use the same values as in Ref.\ 
\cite{frpa}. We evaluate $\delta E$ both quantum-mechanically using the 
Strutinsky method, and semiclassically by the perturbative trace formula 
discussed in Ref.\ \cite{peter}. 

The quantum-mechanical energy spectrum $\{E_i(def)\}$ of the spheroidal 
cavities with the shapes defined by (\ref{multipol2}) was obtained by a 
method developed in Ref.\ \cite{koizu}. In terms of the $E_i(def)$, the 
exact (quantum-mechanical) shell-correction energy is then defined as 
usual \cite{strut} by
\be
\delta E_{qm}= \sum_{i=1}^N E_i 
             - \int_0^{\widetilde\lambda}E\,{\widetilde g}(E)\,dE\,,
\ee
whereby ${\widetilde g}(E)$ and $\widetilde{\lambda}$ are given further
below.

We refer to Ref.\ \cite{peter} for the definition of the modulation 
factor included in the perturbative semiclassical trace formula for the 
oscillating part $\delta g_{sc}(E)$ of the level density. When admitting 
several types of deformations $\eps_j$ simultaneously, the corresponding 
first-order changes of the classical actions $\Delta S_j$ have to be 
added up in the exponent of the integrand for the modulation factor. The 
semiclassical shell-correction energy $\delta E_{sc}$ is then obtained 
using the relation
\be
\delta E_{sc}=\int_0^{\lambda}\!E\,[{\widetilde g}(E)+\delta g_{sc}(E)]\,dE
             -\int_0^{\widetilde\lambda}\!E\,{\widetilde g}(E)\,dE\,,
\ee
where the Fermi energies $\lambda$ and $\widetilde\lambda$ are
determined iteratively by conserving the particle number: 
\be
N = \int_0^{\lambda}[{\widetilde g}(E)+\delta g_{sc}(E)]\,dE
  = \int_0^{\widetilde\lambda}{\widetilde g}(E)\,dE\,.
\label{Nlam}
\ee
Here ${\widetilde g}(E)$ is the average level density found analytically
from its Weyl expansion (see, e.g., Ref.\ \cite{book}) or by a numerical
Strutinsky averaging \cite{strut} of the spectrum $\{E_i\}$.

\Figurebb{du}{60}{40}{795}{490}{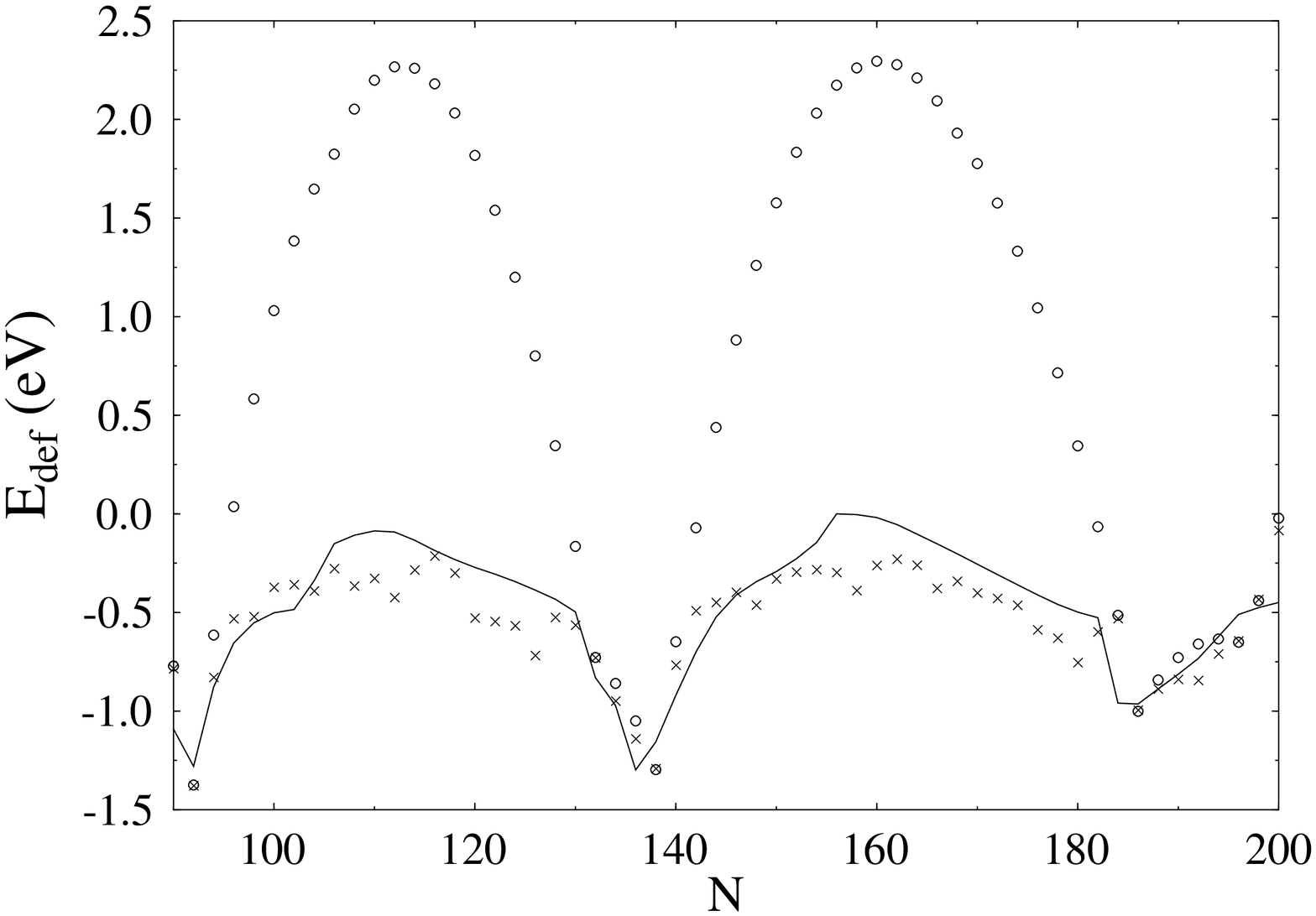}{5.85}{8.5}{
Deformation energy versus atom num\-ber $N$. {\it Circles:} 
quantum-mechanical with spherical shapes; {\it crosses:} 
quantum-mechanical, deformed. {\it Solid line:} semiclassical. For the 
latter two, the energy was minimized for each $N$ with respect to the 
three multipole deformations $\eps_2$,  $\eps_3$, and $\eps_4$.
}

\vspace*{-0.25cm}

In Fig.\ \ref{du} we present deformation energies $E_{\rm def}(def) = 
E_{tot}(def)-E_{LDM}(0)$ versus the particle number $N$ in the region of 
$90\le N \le 200$. The circles show the quantum-mechanical Strutinsky 
results for spherical shapes and exhibit three typical shell closures at 
the magic numbers $N=92$, 138, and 186. The crosses show the quantum
results obtained with deformed shapes, minimizing $E_{tot}$ for each
$N$ with respect to the three multipole deformations $\eps_2$, $\eps_3$,
and $\eps_4$. The solid line is the result of the semiclassical calculation 
including periodic orbits with up to $t_m=3$ repetitions and $p_m=30$ 
reflections. We see how the inclusion of deformations reduces the total 
energy drastically, except for the narrow regions around the magic numbers. 
We obtain a very good agreement of our semiclassical results with the 
quantum results, showing that the perturbative trace formula works 
reasonably well even for the largest deformations.

\Figurebb{defconv}{60}{35}{568}{730}{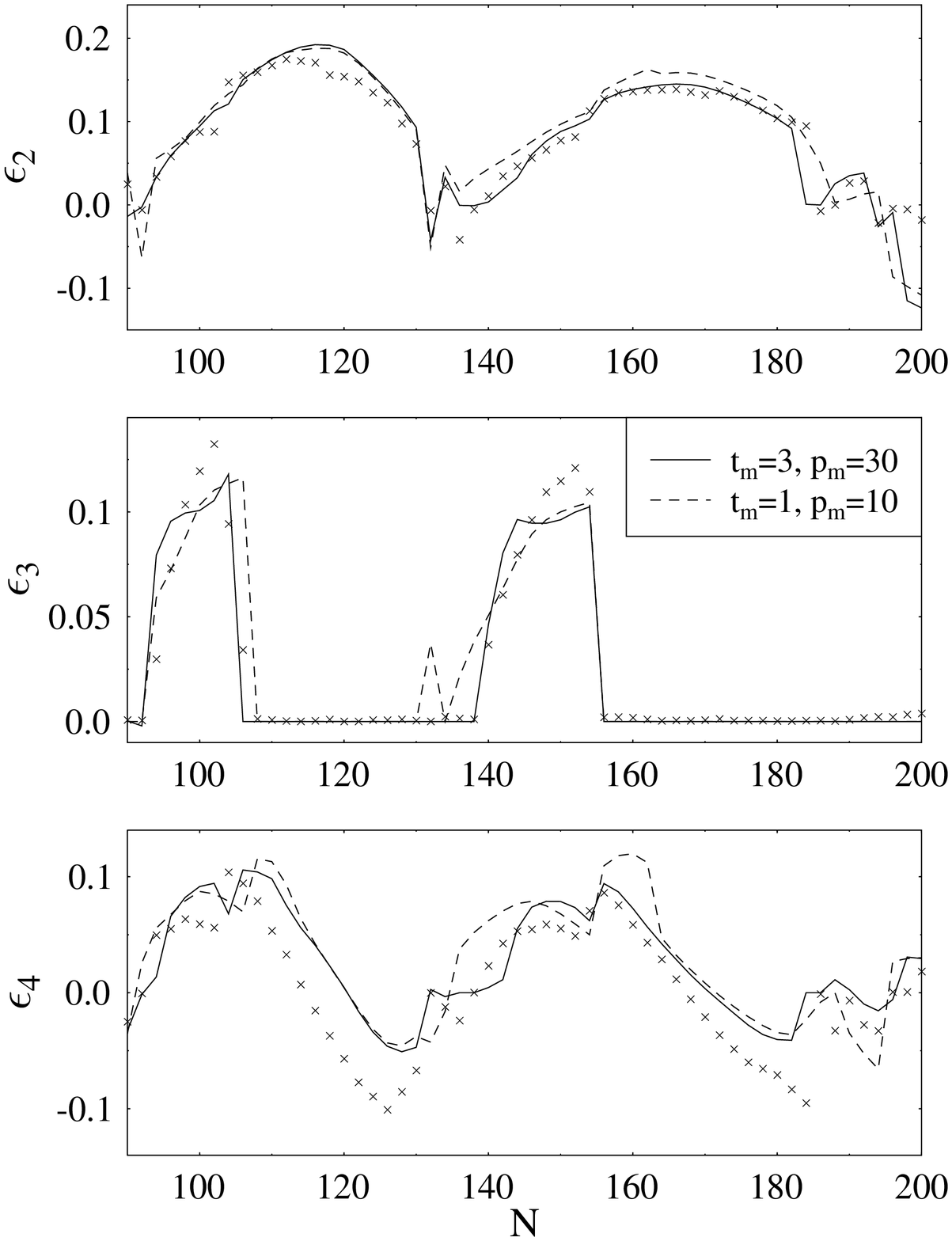}{10.4}{8.5}{
Ground-state quadrupole ($\eps_2$), octupole ($\eps_3$) and
hexadecapole ($\eps_4$) deformations of Na clusters versus number
of atoms $N$. {\it Crosses:} quantum-mechanical results. Lines give
the semiclassical results with different numbers of included periodic 
orbits. {\it Solid lines:} $t_m=3$, $p_m=30$; {\it dashed lines:} 
$t_m=1$, $p_m=10$ (see text). 
}

\vspace*{-0.25cm}

In Fig.\ \ref{defconv}, we present the three ground-state deformations
$\eps_2$, $\eps_3$, and $\eps_4$, obtained by the energy minimization
procedure, as functions of $N$. The crosses are the quantum-mechanical
results, and the solid and dashed lines are the semiclassical results. 
The agreement between semiclassics and quantum mechanics is very 
satisfactory in view of the fact 
that we cannot expect the semiclassical approximation to be exact. In 
particular, even the largest deformations agree within about 10\% - 20\%, 
in spite of the fact that for these deformations the argument of the 
modulation factor in the trace formula is larger than unity. A 
particularly nice result is the correct reproduction of the systematics 
of octupole deformations $\eps_3$ which only occur in the lowest part of 
each shell, as already noticed in Ref.\ \cite{frpa}. In addition, the 
systematic sign change in the hexadecapole deformations $\eps_4$ from 
positive to negative within each shell is reproduced qualitatively, 
although the negative values turn out to be systematically too small in 
the semiclassical results. This defect may be connected to the fact that 
for larger negative hexadecapole deformations the cluster shape 
becomes a multi-valued function in the cylindrical coordinates. Some
minor changes of the deformation systematics are expected when
non-axial shapes are used \cite{triax}, although they can hardly be 
confirmed experimentally.

The convergence of the semiclassical results with respect to the lengths 
of the included periodic orbits is demonstrated by the dashed line in
\fig{defconv}, for which $(t_m,p_m)$ = (1,10) was used, whereas the solid 
line was obtained for $(t_m,p_m)$ = (3,30) as in \fig{du}. We clearly see 
that a slight improvement is obtained by including the second ($t=2$) and 
third $(t=3)$ repetitions (harmonics) and the corresponding numbers of 
reflections.

With the same choice of maximum orbit lengths, we have calculated the
deformation energies and ground-state deformations of clusters with up to 
$N=550$ atoms. The results are presented in Figs.\ \ref{dularge} and 
\ref{deflarge}. Here the same systematics are found as in Fig.\ 
\ref{defconv} above, although they become less distinct for the larger 
clusters where the supershells start to decrease in amplitude.

\Figurebb{dularge}{60}{40}{795}{490}{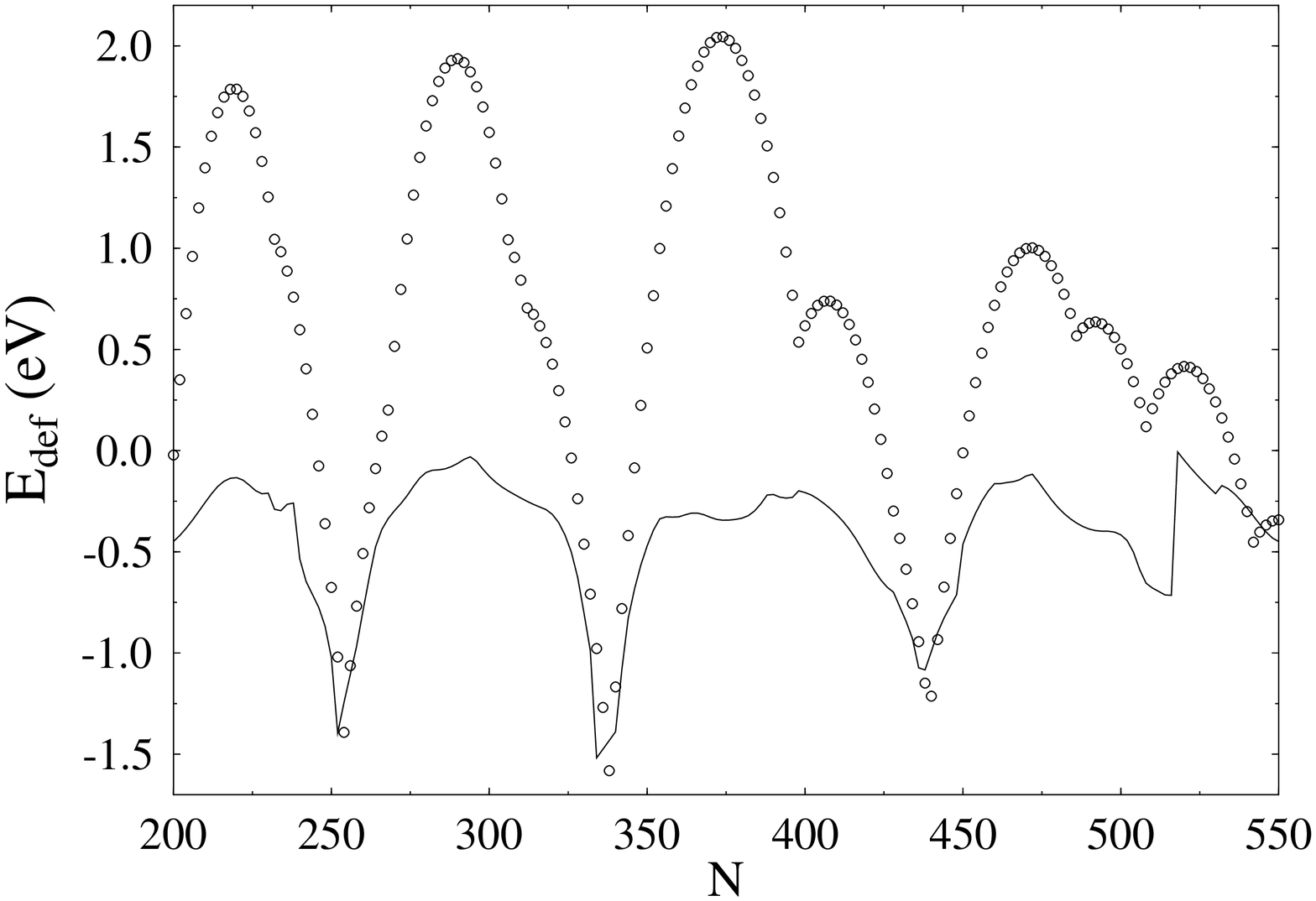}{5.85}{8.5}{
Deformation energy (in eV) versus atom number $N$. {\it Crosses:} 
quantum-mechanical with spherical shapes; {\it solid line:} 
optimized semiclassical results ($t_m=3,$ $p_m=30$).
}

\vspace*{-0.25cm}

Owing to the efficiency of our method, the numerical calculation of 
multidimensional deformation energies becomes quite economical. As 
examples, we show in Figs.\ \ref{surf162} and \ref{surf180} two-dimensional 
deformation energy surfaces of Na$_{162}$ and Na$_{180}$. The 
deformation energy $E_{def}$ is plotted versus quadrupole ($\eps_2$) 
and hexadecapole ($\eps_4$) deformations. The upper parts give the 
quantum-mechanical results and the lower parts the semiclassical ones. We 
see that our semiclassical method captures the correct overall behaviour of 
the topology of the quantum-mechanical deformation energy landscape, thus 
being able to reproduce the most prominent isomeric minima correctly.

We have purposely limited our investigations to cluster sizes $N\simg 90$.
Several reasons limit our approach to such particle numbers. First, 
the selfconsistent total potential (which includes the Coulomb 
repulsion between the electrons) does not have very steep walls for 
particle numbers $N\siml 60 - 80$ (cf.\ \cite{reim2,nish,genz}), so 
that the cavity model cannot be justified. Second, the ionic structure 
effects become more important for smaller clusters, so that any smooth 
shell-model potential becomes a bad approximation. Third, the absolute 
values of the ground-state deformations tend to increase with decreasing 
particle number \cite{reim2}, so that the perturbative approach will break 
down for small clusters.

\Figurebb{deflarge}{105}{0}{580}{710}{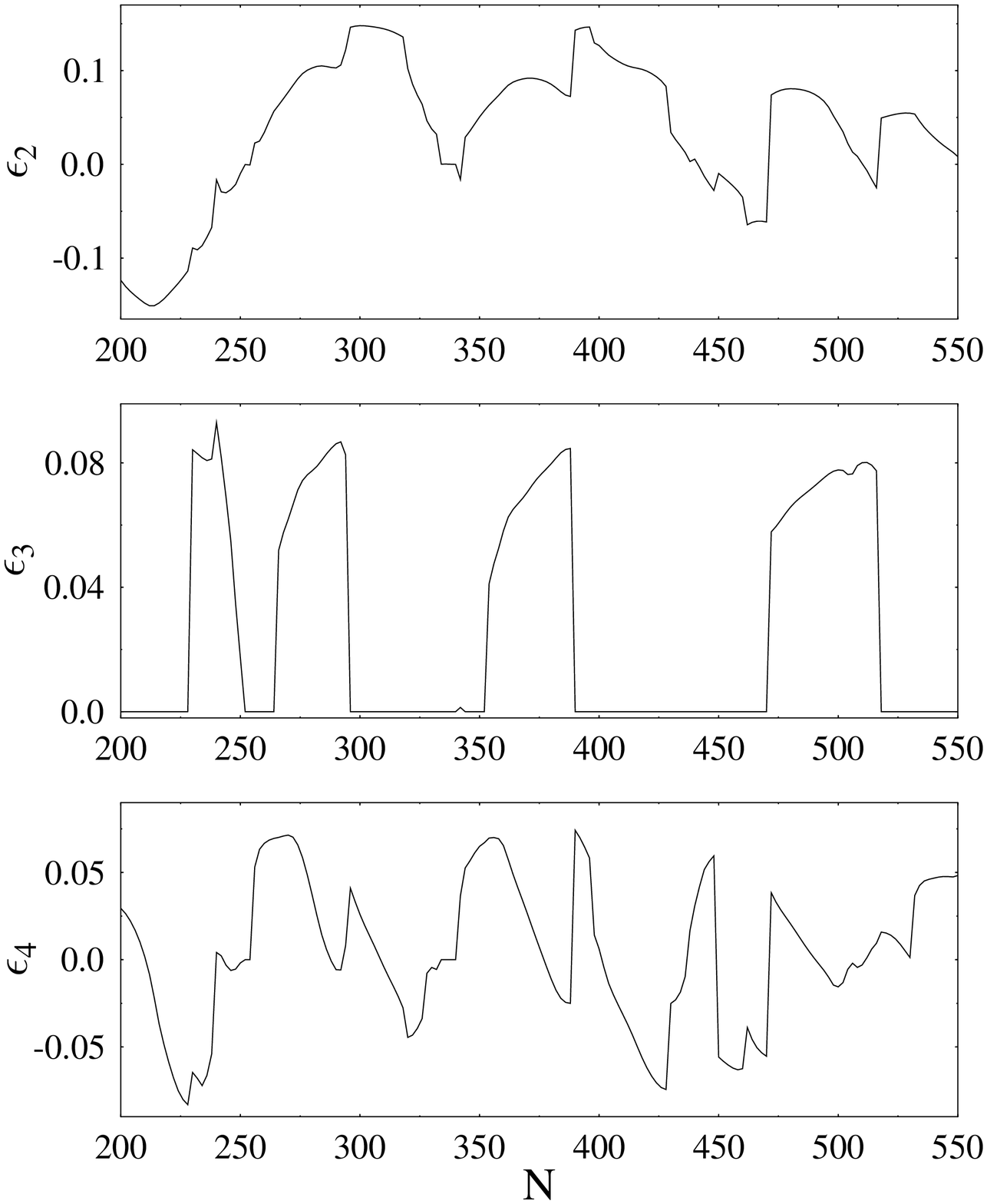}{10.7}{8.5}{
Ground-state deformations for larger clusters with
$200\leq N\leq 550$, obtained semiclassically ($t_m=3,$ $p_m=30$).
}

\vspace*{-0.25cm}

In summary, we have shown that the perturbative semiclassical POT
is an efficient and reliable tool for the calculation of ground-state
deformations of metal clusters, or of any other system of fermions
whose mean field can be approximated by a cavity with reflecting
walls. This holds also for atomic nuclei \cite{strma}, apart from 
modifications due to the spin-orbit interaction whose inclusion in 
the POT is still an object of actual research \cite{bolke,luso}. 
Our method can easily be extended to describe charged clusters by
using the relevant liquid drop model developed, e.g., in Ref.\ 
\cite{seidl}.

\vspace*{0.2cm}

We are grateful to S. Frauendorf for his interest and encouraging
discussions at early stages of this work. P.M. 

\Figurebb{surf162}{25}{15}{468}{280}{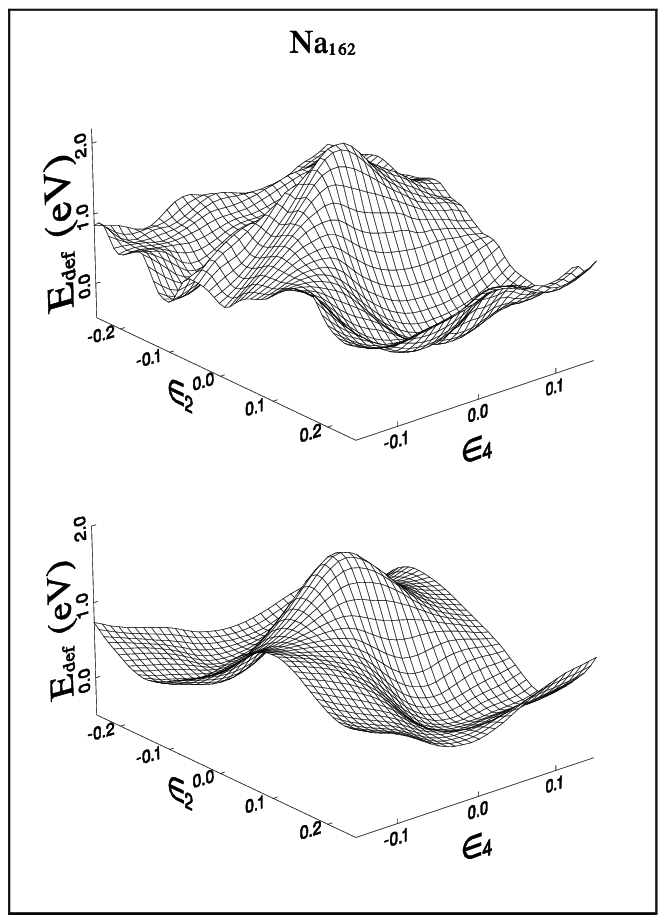}{11.3}{8.5}{
Deformation energy $E_{def}$ of Na$_{162}$ versus deformations 
$\eps_2$ and $\eps_4$. {\it Upper part:} quantum-mechanical, 
{\it lower part:} semiclassical result ($t_m=3,$ $p_m=30$). 
}

\vspace*{-0.25cm}

\noindent
acknowledges the warm 
hospitality and support of the Bogolyubov Laboratory at Dubna. 
This work was partially supported by INTAS, the Landau-Heisenberg 
program, and the Deutsche Forschungsgemeinschaft.

\end{document}